\documentclass[conference,10pt]{IEEEtran}

\usepackage{graphicx} 
\usepackage{hyperref}
\urldef{\myhrefurl}\url{https://github.com/ismail0T/Strategic_Bidding_P2P_Energy_Trading}

\usepackage{cite}
\usepackage{amsmath,amssymb,amsfonts}
\usepackage{algorithmic}
\usepackage{graphicx}
\usepackage{textcomp}
\usepackage{xcolor}
\def\BibTeX{{\rm B\kern-.05em{\sc i\kern-.025em b}\kern-.08em
    T\kern-.1667em\lower.7ex\hbox{E}\kern-.125emX}}
\usepackage{graphicx}
\usepackage{subcaption}
\graphicspath{ {images/} }
\usepackage[utf8]{inputenc}
\usepackage{fancyhdr}
\usepackage{dirtytalk}
\usepackage{tabularx}
\usepackage[ruled,vlined]{algorithm2e}

\usepackage{enumitem,kantlipsum}
\usepackage{listings}
\usepackage{graphicx}
\graphicspath{ {images/} }
\usepackage[utf8]{inputenc}
\usepackage{fancyhdr}
\usepackage{dirtytalk}
\usepackage{tikz}
\usetikzlibrary{positioning,arrows.meta}
\usepackage{bbm}
\usepackage{booktabs}
\usetikzlibrary{shapes.geometric, arrows}
\usepackage{pgfplots}
\pgfplotsset{width=8cm,compat=1.16}
\usepgfplotslibrary{external}
\usepackage{array}
\newcolumntype{P}[1]{>{\centering\arraybackslash}p{#1}}

\usepackage[most]{tcolorbox}

\tikzstyle{startstop} = [rectangle, rounded corners, minimum width=2cm, minimum height=0.5cm,text centered, draw=black, fill=red!30]
\tikzstyle{process} = [rectangle, minimum width=2cm, minimum height=0.5cm, text centered, draw=black, fill=orange!30, align=left]
\tikzstyle{decision} = [diamond, minimum width=1.0cm, minimum height=0.4cm, text centered, draw=black, fill=green!30]
\tikzstyle{arrow} = [thick,->,>=stealth]

\newcolumntype{L}{>{$}l<{$}}

\usepackage[labelformat=simple]{subcaption}

\usepackage[nolist]{acronym}
\begin{acronym}
\acro{LM}{language model}
\acro{LLM}{large language model}
\acro{SLM}{small language model}
\acro{RAG}{retrieval augmented generation}
\acro{AI}{artificial intelligence}

\acro{MDP}{Markov decision process}
\acro{IC}{incentive compatibility}
\acro{IR}{individual rationality}
\acro{URLLC}{ultra reliable low latency}
\acro{AI}{artificial intelligence}

\acro{VCG}{Vickrey–Clarke–Groves}
\acro{GFP}{Generalized First Price}
\acro{GSP}{Generalized Second Price}
\acro{UCB}{upper confidence bound}

\acro{ZIP}{zero-intelligent plus}
\acro{CDA}{continuous double auction}
\acro{MAB}{multi-armed bandit}
\acro{UCB}{upper confidence bound}
\acro{RL}{reinforcement learning}
\acro{MADDPG}{multi-agent deep deterministic policy gradient}
\acro{DDPG}{deep deterministic policy gradient}
\acro{PPO}{proximal policy optimization}
\acro{DSO}{distribution system operator}
\acro{P2P}{peer-to-peer}
\acro{HEMS}{home energy management systems}
\acro{MARL}{multi-agent reinforcement learning}
\acro{DER}{distributed energy resource}
\acro{FIT}{feed-in-tariff}
\acro{POMDP}{partially observable Markov decision process}
\acro{CTDE}{centralized-training, decentralized-execution}

\end{acronym}

\begin{document}
\captionsetup[figure]{labelfont={bf},labelformat={default},labelsep=period,name={Figure}}

\title{Large Language Models as Strategic Bidding Agents in P2P Energy Trading Markets}



\author{\IEEEauthorblockN{Ismail Lotfi\IEEEauthorrefmark{1}, Ali Ghrayeb\IEEEauthorrefmark{1} and Haitham Abu-Rub\IEEEauthorrefmark{1}}

\IEEEauthorblockA{\IEEEauthorrefmark{1}College of Science and Engineering, Hamad Bin Khalifa University, Doha, Qatar}

}

\maketitle

\begin{abstract}
Peer-to-peer (P2P) energy trading markets rely on double auction mechanisms to match prosumers and consumers in smart grid distribution networks. However, the strategic behavior of bidding agents in such markets remains not fully explored, particularly in repeated settings with bounded rationality. This paper proposes a novel framework that integrates large language models (LLMs) as reasoning-driven bidding agents in repeated P2P energy double auctions. We compare the performance of three bidding strategies: random bidding, an $\varepsilon$-greedy multi-armed bandit (MAB) approach, and an LLM-based strategy.
Simulation results show that the LLM-based strategy achieves superior cleared trading volume over the first episodes compared to $\varepsilon$-greedy MAB and random bidding baselines, eliminating the exploration burn-in period that statistical learning algorithms inherently require before converging to productive price arms. 
Importantly, when tested in an environment different from the one used during learning, the performance of the $\varepsilon$-greedy strategy drops significantly, while that of the LLM-based bidding strategy continue to achieve higher surplus and successful trades.
Nevertheless, the LLM's advanced contextual reasoning also gives rise to an important market dynamic. In a homogeneous population of LLM agents, sellers increasingly exploit buyers' rational outside options to drive clearing prices above the Nash equilibrium, resulting in a progressively more asymmetric allocation of surplus in favor of sellers that does not converge within the observed time horizon.
\end{abstract}


\section{Introduction}

\subsection{Background}

The rapid deployment of \acp{DER}, such as rooftop photovoltaic systems and small wind turbines, has transformed electricity consumers in smart grids into \emph{prosumers} capable of both producing and consuming energy~\cite{Tushar2020}. As traditional compensation schemes, including \ac{FIT} and net metering, face increasing scalability and equity challenges as grid infrastructure costs are disproportionately borne by non-DER consumers~\cite{Zibo_2023_TSG}. \Ac{P2P} energy trading has emerged as a promising alternative that enables direct energy exchange among participants without relying exclusively on the utility.

Double auction mechanisms are particularly well suited for P2P energy trading in smart grids because they efficiently match buyers and sellers while balancing traded volume, budget balance, and surplus allocation~\cite{Zibo_2023_TSG,Dawei_2023_TPWRS,pereira2022peer}. However, their performance strongly depends on the bidding strategies adopted by market participants. Designing intelligent bidding agents is especially challenging in repeated auctions, where agents operate under bounded rationality and must continuously adapt to changing market conditions.

\subsection{Related Works}

Early studies on P2P energy trading relied on heuristic bidding strategies or statistical learning methods. Guerrero \textit{et al.}~\cite{Guerrero_2019_TSG} employed \ac{ZIP} agents within a continuous double auction, while Zhao \textit{et al.}~\cite{Zibo_2023_TSG} investigated \ac{MAB} algorithms, including \ac{UCB} and $\varepsilon$-greedy, in large-scale P2P markets. Owing to their decentralized nature, MAB methods scale efficiently to thousands of participants. However, they eventually converge to nearly static bidding policies and must re-explore whenever market conditions change.
To improve adaptability, several studies have adopted \ac{RL}, particularly \ac{MARL}, to learn bidding policies through repeated market interactions~\cite{Yujian_2020_TSG,qiu2021_IJCAI,Dawei_2023_TPWRS,Jiehui_2024_TII}. Although these approaches can capture strategic interactions among agents, they typically require centralized training, substantial computational resources, and relatively stable environments, which limits their scalability and practical deployment.

More recently, \acp{LLM} have demonstrated strong reasoning and decision-making capabilities in strategic environments, including auction-based resource allocation~\cite{mao_2024_Alympics_LLM,Ismail_2026_LLM_WCNC}. In smart grids, however, LLMs have primarily been used as auxiliary modules for reward shaping, market assessment, or price forecasting rather than autonomous bidding agents~\cite{Qianyi_2025_TEMPR,Jadhav2026,Chengwei_2026_TSG,Xin_2025_TEMPR}. Consequently, their ability to directly participate in repeated P2P energy double auctions remains largely unexplored.

\subsection{Research Gap and Contributions}

Learning-based methods, including MAB and \ac{RL}, optimize their policies for the environment encountered during training. However, when the underlying environment changes, the learned policy may no longer remain optimal, requiring additional exploration or even complete retraining before recovering its performance. This limitation is particularly relevant in P2P energy markets, where the market conditions are inherently dynamic. Changes in the number of participating buyers and sellers, pricing rules, renewable generation, electricity demand, or user behavior can all alter the reward distribution experienced by learning agents. In contrast, LLMs rely on contextual reasoning rather than memorized policies, allowing them to adapt their bidding decisions without retraining. 
However, it is unclear whether an LLM can replace conventional bidding algorithms and autonomously participate in repeated P2P energy markets. Furthermore, little is known about the market dynamics that emerge when multiple reasoning-capable LLM agents repeatedly interact, particularly regarding market efficiency, surplus allocation, and strategic behavior. Understanding these interactions is becoming increasingly important given recent evidence that LLM agents may exhibit sophisticated strategic or even collusive behaviors in auction settings~\cite{agrawal2025_LLM_collusion_auction,lin2024strategiccollusionllmagents}.
To address these questions, this paper proposes an LLM-driven bidding framework for repeated P2P energy double auctions and evaluates it against conventional decentralized bidding strategies. The main contributions are summarized as follows:

\begin{enumerate}
    \item \textbf{Autonomous LLM bidding framework.} We propose, to the best of our knowledge, the first framework that employs an LLM as a standalone bidding agent for double auctions in P2P energy trading. Unlike existing approaches that use LLMs only as auxiliary modules, the proposed agent generates bid prices directly from observable market information through natural-language reasoning, without policy optimization.

    \item \textbf{Comprehensive benchmarking.} We compare the proposed LLM bidding strategy with random bidding and the widely adopted $\varepsilon$-greedy MAB algorithm under the $k$-double auction mechanism~\cite{Zibo_2023_TSG}. The comparison evaluates cleared trading volume, auction surplus, convergence characteristics, and bidding behavior over repeated market interactions. Simulation results demonstrate that the LLM-based strategy outperforms both baselines in terms of market efficiency and bid precision, and adapts immediately to changing of markets dynamics without requiring a relearning period, a limitation that $\varepsilon$-greedy and other learning-based methods cannot overcome by design.

    \item \textbf{Characterization of emergent market behavior.} We show that homogeneous populations of LLM agents (i.e., all agents uses LLM-based bidding strategy) exhibit qualitatively different strategic behavior, where sellers progressively exploit buyers' outside options to increase clearing prices beyond the Nash equilibrium, producing an increasingly asymmetric surplus distribution. These findings reveal both the potential and the risks of deploying reasoning-capable AI agents in future electricity markets.
\end{enumerate}

The remainder of this paper is organized as follows. Section~II presents the system model, auction mechanism and the bidding strategies. Section~III presents the LLM-based biding strategy and addresses some practical deployment challenges. Section~IV presents the numerical results. Section~V concludes the paper.

\section{System Model and Baseline Bidding Strategies}

In this section, we describe our studied system model, formalize the $k$-Double auction mechanism and then describe the baseline bidding strategies employed by bidding agents.

\subsection{Market Participants and Roles}
We consider a local P2P energy distribution network within the smart grid ecosystem, comprising two categories of agents: pure consumers (buyers) and prosumers. Prosumers are \ac{DER} owners (e.g., households equipped with rooftop photovoltaic systems) that participate in the market by selling their excess renewable generation. 
Following~\cite{Zibo_2023_TSG}, we assume that sellers (prosumers with excess PV generation) have zero marginal costs, and all DER quantities not consumed locally are offered to the market (i.e., no local storage).

\subsection{Auction Surplus and Learning}

The performance of a P2P energy market is evaluated using the \emph{auction surplus}, which measures the economic benefit obtained from trading relative to the utility as the outside option. Buyers can always purchase electricity from the grid at the retail price $P_{\text{UR}}$, while sellers can sell surplus energy to the grid at the feed-in tariff $P_{\text{FIT}}$, where $P_{\text{FIT}} < P_{\text{UR}}$. Consequently, any clearing price $p_i^{\text{au}} \in (P_{\text{FIT}}, P_{\text{UR}})$ benefits both parties.
The surplus (reward) of agent $i$ is then defined as:
\begin{equation}
  \bar{S}_i =
  \begin{cases}
    (P_{\text{UR}} - p_i^{\text{au}}) q_i^{\text{au}}, & \text{buyer},\\
    (p_i^{\text{au}} - P_{\text{FIT}}) q_i^{\text{au}}, & \text{seller},
  \end{cases}
  \label{eq:auc_surplus}
\end{equation}
where $q_i^{\text{au}}$ is the cleared energy quantity.

To facilitate learning, rewards are normalized to $[0,1]$:
\begin{equation}
    \pi_{i,b}^{t} =
    \frac{P_{\text{UR}} - p_i^{\text{au},t}}
         {P_{\text{UR}} - P_{\text{FIT}}}, \hspace{0.2cm} \text{buyer},
    \label{eq:price_norm_buyer}
\end{equation}

\begin{equation}
    \pi_{i,s}^{t} =
    \frac{p_i^{\text{au},t} - P_{\text{FIT}}}
         {P_{\text{UR}} - P_{\text{FIT}}}, \hspace{0.2cm} \text{seller}.
    \label{eq:price_norm_sell}
\end{equation}

If an agent is not matched in auction round $t$ (i.e., $q_i^{\text{au},t}=0$), its normalized reward is set to zero, corresponding to the utility fallback outcome.

\subsection{$k$-Double Auction Mechanism}\label{sec:k_double_auction_description}

In the $k$-Double auction~\cite{Zibo_2023_TSG,friedman2018_DA}, buyer bids and seller asks are first sorted in descending and ascending order, respectively:
\begin{equation}
    pb_1 > pb_2 > \cdots > pb_m,\qquad
    ps_1 < ps_2 < \cdots < ps_n.
\end{equation}

The intersection of the demand and supply curves determines the traded quantity $Q^*$, with marginal buyer bid $p_{bL}$ and marginal seller ask $p_{sH}$. The clearing price is computed as
\begin{equation}\label{eq:clearing_price}
P^* = k\,p_{bL} + (1-k)\,p_{sH}, \qquad k\in[0,1].
\end{equation}
All matched agents trade at $P^*$, while unmatched agents trade with the grid.
In the repeated auction setting, each agent selects a bid at every round without knowing future market outcomes, giving rise to an exploration--exploitation trade-off. The following subsections compare three bidding strategies with increasing sophistication: random bidding, $\varepsilon$-greedy multi-armed bandits, and the proposed LLM-based bidding agent.

\subsection{Baseline Bidding Strategies}
\subsubsection{Random Bidding}
Buyers and sellers bid random values in the range $[P_\text{FIT}, P_\text{UR}]$~\cite{Dawei_2023_TPWRS}.

\subsubsection{$\varepsilon$-Greedy Algorithm}

The \ac{MAB} framework models each bid price as an arm with an unknown reward distribution. At round $t$, agent $i$ maintains, for each arm $k$, the number of selections $n_i(t,k)$ and its empirical average normalized reward~\cite{Zibo_2023_TSG,kuleshov2014}:
\begin{equation}
  \bar{\pi}_i(k)
  =
  \frac{1}{n_i(t,k)}
  \sum_{\tau=1}^{t}
  \pi_i^{(\tau)}
  \mathbf{1}\{\xi^{(\tau)}=k\},
  \label{eq:avg_reward}
\end{equation}
where $\xi^{(\tau)}$ is the selected arm and $\pi_i^{(\tau)}$ is the normalized reward.
The $\varepsilon$-greedy policy balances exploration and exploitation by selecting the arm with the highest empirical reward with probability $1-\varepsilon$, and a uniformly random arm with probability $\varepsilon$:
\begin{equation}
  \xi^{(t)}
  =
  \begin{cases}
    \arg\max_k \bar{\pi}_i(k), & \text{with probability } 1-\varepsilon,\\
    \text{Uniform}\{1,\ldots,K\}, & \text{with probability } \varepsilon.
  \end{cases}
  \label{eq:eps_greedy}
\end{equation}

\section{Methodology}
\subsection{LLM-Based Bidding Strategy}
We introduce an LLM-based bidding agent in which the language model reasons over a structured prompt containing:
The agent's current role (buyer/seller) and the agent's historical reward trajectory and the current auction mechanism type.
The prompt template is illustrated in Figure.~\ref{fig:llm_prompt}. The LLM outputs a recommended bid value $b^*$ and a natural-language explanation $\mathcal{E}$ of its reasoning. We use \texttt{gpt-5-mini} for all LLM experiments as used in previous works (e.g.~\cite{Jadhav2026, Ismail_2026_LLM_WCNC}), chosen for its balance of reasoning depth and inference latency.
The LLM is particularly advantageous in this setting because: (i) it can reason about the P2P market changes and adjust bids accordingly; (ii) and it can anticipate bang-bang price dynamics and avoid extreme bids that lead to zero reward.

\begin{figure}[ht]
\centering
\small
\colorbox{gray!10}{
\begin{minipage}{0.95\columnwidth}
\textbf{LLM Prompt Template (P2P Energy Auction)}\\[2pt]
\textbf{Context:}
\begin{itemize}
    \item Role: \texttt{[buyer/seller]}, 
    \item Reserve prices: $P_\text{FIT}$ = \texttt{[\$]}, $P_\text{UR}$ = \texttt{[\$]}
    \item Last bidding prices and rewards (last 10 rounds): \texttt{[value]}
    \item Mechanism: \texttt{[k-double]}
\end{itemize}
\textbf{Task:} Recommend a bid price to maximize your surplus over future rounds. Justify your reasoning briefly.\\[2pt]
\textbf{Response format:}\\
\texttt{Bid: [value]} \\
\texttt{Reasoning: [explanation]}
\end{minipage}
}
\caption{LLM prompt template for P2P auction bidding.}
\label{fig:llm_prompt}
\end{figure}

\subsection{Practical Deployment Challenges}

A fundamental tension inherent to deploying \acp{LLM} as bidding agents in \ac{P2P} energy markets is that the most capable \ac{LLM} reasoning engines are cloud-hosted and require sensitive prosumer data to be transmitted off-premises, while privacy-preserving local deployment faces severe computational constraints on residential \ac{HEMS} hardware. We highlight bellow two primary approaches that can be used to alleviate this challenge:
\begin{itemize}
    \item \textit{Differentially Private Prompt Construction:} Differential privacy~\cite{Demelius2025_DP} provides a rigorous mathematical framework for quantifying and bounding the privacy leakage of data-dependent computations.
    In the \ac{LLM} bidding context, differential privacy can be applied at the prompt construction stage: before any numerical feature is included in the \ac{LLM} prompt, calibrated Gaussian or Laplace noise is added to each sensitive quantity.
    For a feature $x$ with global sensitivity $\Delta x$ (the maximum change in $x$ due to a single agent's data), the Gaussian mechanism adds noise $\eta \sim \mathcal{N}(0, \sigma^2)$ with $\sigma = \Delta x \cdot \sqrt{2 \ln(1.25/\delta)} /\varepsilon_\text{priv}$ to produce the privatized feature $\tilde{x} = x + \eta$ included in the prompt, where $\delta  \in (0,1)$ is the failure probability and $\varepsilon_\text{priv}$ is the privacy budget.

    \item \textit{Edge-Deployable Small Language Models:} The most structurally clean solution is to eliminate the cloud dependency entirely by deploying a sufficiently compact \ac{LLM} directly on the \ac{HEMS} device. Recent advances in model compression, including quantization, pruning, and knowledge distillation, have produced a new class of \emph{small language models} (SLMs) that retain substantial reasoning capability at a fraction of the computational cost of frontier models~\cite{Wang2025}.
    Modern \ac{HEMS} devices and smart home hubs (e.g., those based on Raspberry Pi 5, NVIDIA Jetson Nano) already meet the memory and compute requirements for SLM parameter models at 4-bit quantization~\cite{Ardakani_2025_CVPR}. However, the monetary and reasoning quality trade-offs between deploying LLMs on the edge or on the cloud should be investigated.

\end{itemize}

\section{Numerical Results}\label{sec_results} 
We evaluate the performance of LLM-based, $\varepsilon$-greedy, and random bidding agents through simulations of a repeated k-double P2P energy auction with $k=0.5$\footnote{Code available at: \myhrefurl}. Each simulation consists of $D=100$ auction rounds, representing daily settlement periods. Without loss of generality, the feed-in tariff and utility rate are set to $P_{\text{FIT}}=10$\,\textcent/kWh and $P_{\text{UR}}=20$\,\textcent/kWh, respectively. Performance is evaluated using two primary metrics: the cleared trading volume $\bar{Q}^{h}$ and the agents' surplus $\bar{S}^t$, reported separately for buyers and sellers to capture the distributional effects of each bidding strategy. To isolate and better analyze market behavior on each side of the auction, prosumers are assumed to act exclusively as sellers throughout the simulation, without switching between buying and selling roles. For each of the three bidding strategies, we consider the market consists of $M=10$ consumers (buyers) and $N=10$ prosumers (sellers only), and that supply/demand quantities are balanced\footnote{Modeling time-varying demand and supply patterns across daily and seasonal horizons is left for future work.}.

\subsection{Cleared Trading Volume}

\begin{figure}[ht]
    \centering
    \includegraphics[width=0.99\columnwidth,height=5.1cm]{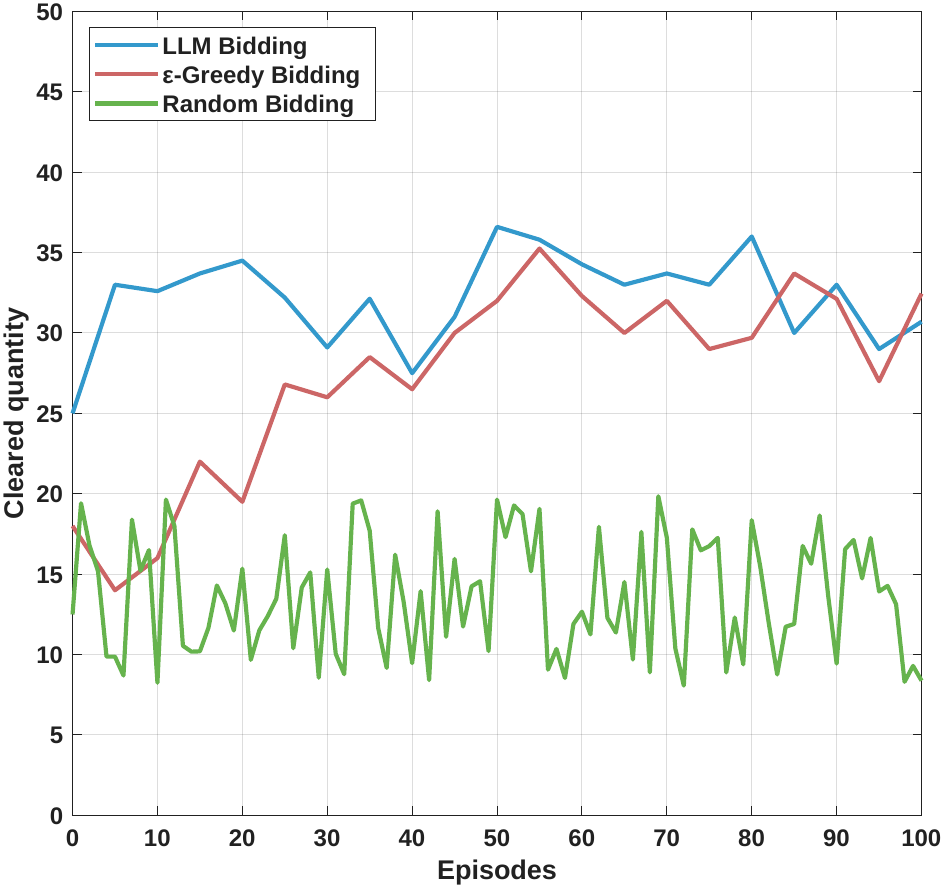}
    \caption{Average cleared trading volume $\bar{Q}^{h}$
    over episodes for LLM-based, $\varepsilon$-greedy and random
    bidding agents in the k-double P2P energy auction
    ($k = 0.5$).}
    \label{fig:cleared_volume}
\end{figure}

Figure.~\ref{fig:cleared_volume} presents the evolution of the average cleared volume $\bar{Q}^{h}$ over episodes for the three bidding strategies.
Two distinct behavioral phases emerge from Figure.~\ref{fig:cleared_volume}. In the
``early phase" (approximately the first 40 episodes), the LLM agent achieves substantially higher cleared volume than both the $\varepsilon$-greedy and random bidding agents. This advantage stems directly from the LLM's ability to reason near-optimally from the very first episode: LLM buyers bid within the clearing zone $(P_{FIT}, P_{UR})$ immediately, and LLM sellers ask within a compatible price range, producing a non-zero intersection of the supply and demand curves from round 1. The $\varepsilon$-greedy agent, by contrast, incurs a significant ``exploration tax" in early episodes as its random arm selection frequently produces extreme bids (very low buyer bids or very high seller asks) that fall outside the clearing zone, resulting in failed trades and reduced $\bar{Q}^{h}$. This early-phase exploration cost corresponds to a real financial penalty in a live P2P market: for approximately $D_{\text{learn}} \approx 40$ calendar days after market launch, $\varepsilon$-greedy households consistently fail to clear trades that LLM-equipped agents would have secured. 

In the ``convergence phase" (after approximately episode 40), the cleared volumes of both strategies converge to a similar level. This is consistent with the theoretical convergence of $\varepsilon$-greedy algorithms~\cite{Zibo_2023_TSG}: after $D_{\text{learn}} \approx 40$ episodes, the algorithm has identified its locally optimal price arms and shifted from exploration to exploitation. Once $\varepsilon$-greedy agents have converged, the frequency of extreme bids diminishes and the supply and demand curves reliably intersect, restoring cleared volume to a level comparable to the LLM strategy. As random bidding agents do not have a built-in mechanism to learn, their clearing volume remains unstable and lower than both LLM and $\varepsilon$-greedy agents for the entire episodes.
These results highlight that the LLM's volume advantage is concentrated in the critical early deployment phase, precisely the period when market liquidity is most fragile and when clearing failures have the greatest impact on participant trust and platform adoption. The LLM effectively eliminates the burn-in period that statistical learning algorithms inherently require, making it immediately deployable on the first day of market operation, which we further emphasis next.

\subsection{Buyers' and Sellers' Surplus}
Figure.~\ref{fig:surplus} presents the evolution of the buyers' surplus $\bar{S}_b^t$ and sellers' surplus $\bar{S}_s^t$ as defined in~\eqref{eq:auc_surplus} over episodes for the three different bidding strategies. 
Tow important patterns emerge from Figure.~\ref{fig:surplus_clearing_price}:



 \begin{figure}[ht] 
     \centering
     \begin{subfigure}[b]{0.48\textwidth}
         \centering         \includegraphics[width=\textwidth,height=5.1cm]{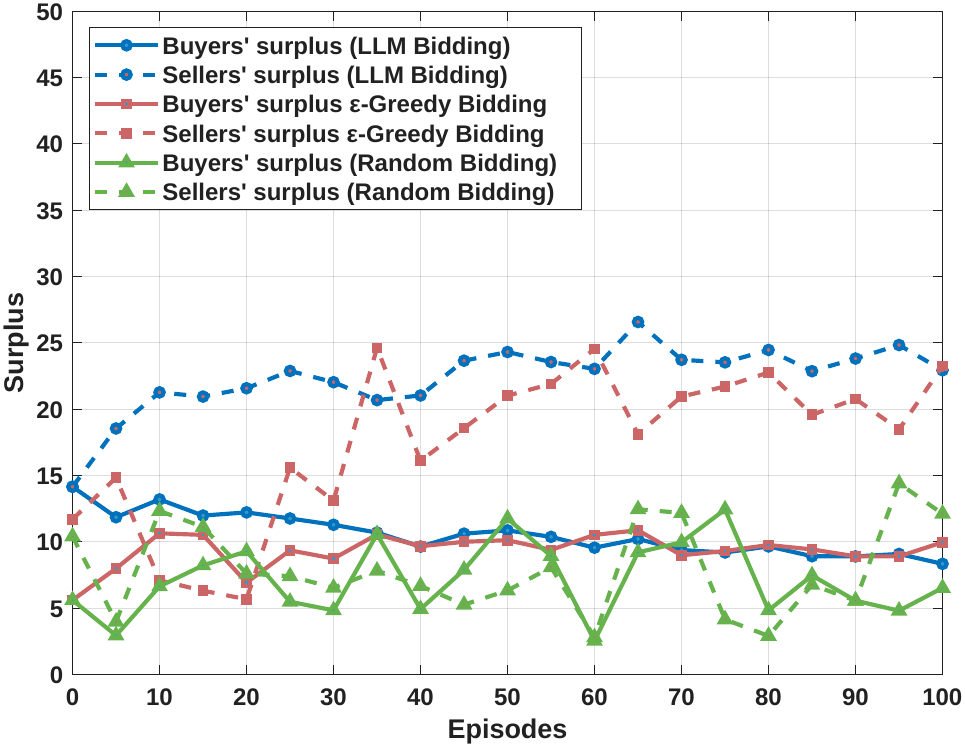}
         \caption{}
         \label{fig:surplus}
     \end{subfigure}
     \begin{subfigure}[b]{0.48\textwidth}
         \centering         \includegraphics[width=\textwidth,height=5.1cm]{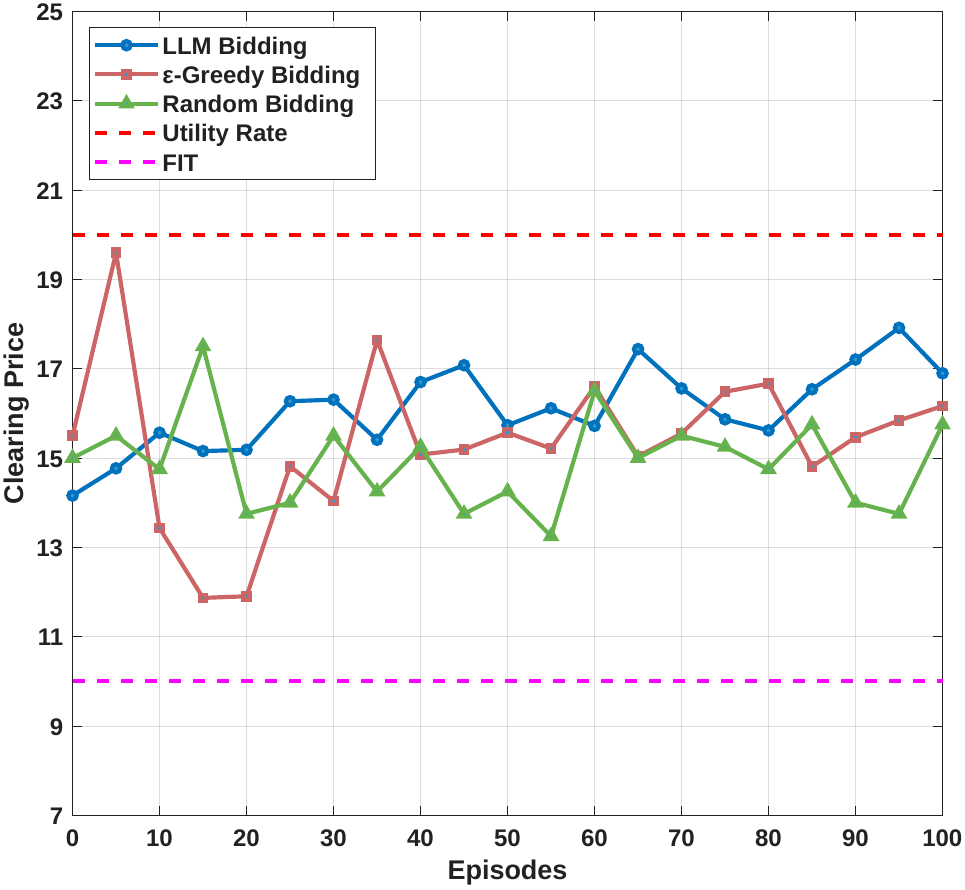}
         \caption{}
         \label{fig:clearing_price}
     \end{subfigure}
     \caption{(a) Average buyers' surplus $\bar{S}_b^t$ and
    sellers' surplus $\bar{S}_s^t$ over episodes for
    LLM-based, $\varepsilon$-greedy and random bidding agents and, (b) clearing price for LLM-based, $\varepsilon$-greedy and random bidding agents.
        }        \label{fig:surplus_clearing_price}
\end{figure}


\paragraph{LLM Surplus Dynamics.}
Under the LLM strategy, Figure.~\ref{fig:surplus} reveals a monotonically increasing seller surplus accompanied by a monotonically decreasing buyer surplus throughout the entire simulation horizon, with the clearing price (Figure.~\ref{fig:clearing_price}) rising continuously from approximately $14$\,\textcent/kWh to $18$\,\textcent/kWh and never stabilizing within the observed horizon. This behavior reflects a systematic and unbounded surplus redistribution from buyers to sellers driven by the LLM's contextual reasoning capability. LLM seller agents observe the clearing price history in their
prompt context at each episode and continuously infer that buyers remain willing to accept higher prices, since a buyer's alternative to clearing at a high price is not clearing at all and paying $P_{\text{UR}} = 20$\,\textcent/kWh to the utility. This rational outside option compels LLM buyers to maintain high bids, which in turn removes the competitive resistance that would otherwise cap seller ask escalation. The result is a self-reinforcing cycle: successful clearing at price $P^{*,t}$ signals to LLM sellers that a higher ask $P^{*,t} + \mu$ (where $\mu$ is a small positive number) is feasible in the next episode, producing continuous upward drift in the clearing price toward $P_{\text{UR}}$ well beyond the theoretical Nash equilibrium of $P^* = \frac{20+10}{2}=15$\,\textcent/kWh~\cite{Zibo_2023_TSG}.

Notably, since cleared volume $\bar{Q}^{t}$ remains high and stable (Figure.~\ref{fig:cleared_volume}), the declining buyer surplus as defined in~\eqref{eq:auc_surplus} is attributable entirely to the rising clearing price rather than any reduction in cleared quantity. LLM buyers rationally accept progressively lower surplus, preferring a diminishing but positive payoff over the zero payoff of a failed trade. This outcome reveals an important limitation of deploying LLM agents in a homogeneous population (i.e., all agents uses LLM-based bidding strategy): their individual rationality collectively enables seller market power to grow unchecked, producing a highly asymmetric market outcome in which sellers capture an increasing share of the available trading surplus at buyers' expense.
Such behavior of LLM agents has been also observed by other related works outside the auction domain, e.g., in~\cite{lin2024strategiccollusionllmagents}, and can be understood through the lens of ``environment non-stationarity". In repeated multi-agent interactions, the environment perceived by each agent is non-stationary because the transition dynamics are determined not only by the agent's own actions but also by the continually evolving strategies of all other participants.

\paragraph{$\varepsilon$-Greedy Surplus Dynamics.}
Under the $\varepsilon$-greedy strategy, Figure.~\ref{fig:surplus} and Figure.~\ref{fig:clearing_price} show a qualitatively different pattern: seller surplus increases over episodes and stabilizes around episode~40, while buyer surplus remains approximately constant throughout. At convergence, the clearing price stabilizes at approximately $16.5$\,\textcent/kWh, above the Nash equilibrium of $15$\,\textcent/kWh but below the LLM's end-of-horizon price of $18$\,\textcent/kWh.
This apparent paradox (i.e., seller surplus increasing without a corresponding decrease in buyer surplus when $\varepsilon$-Greedy is used) is resolved by recognizing that the seller surplus growth is driven primarily by \emph{increasing cleared volume} rather than by rising clearing price. In early episodes, the $\varepsilon$-greedy exploration tax suppresses $\bar{Q}^{h}$, limiting the total surplus available to both buyers and sellers. As the algorithm converges toward episode~40 and cleared volume recovers, additional surplus is generated.

\subsection{Robustness to Post-Deployment Market Changes}
\begin{figure}[ht] 
     \centering
     \begin{subfigure}[b]{0.24\textwidth}
         \centering         \includegraphics[width=\textwidth,height=3.1cm]{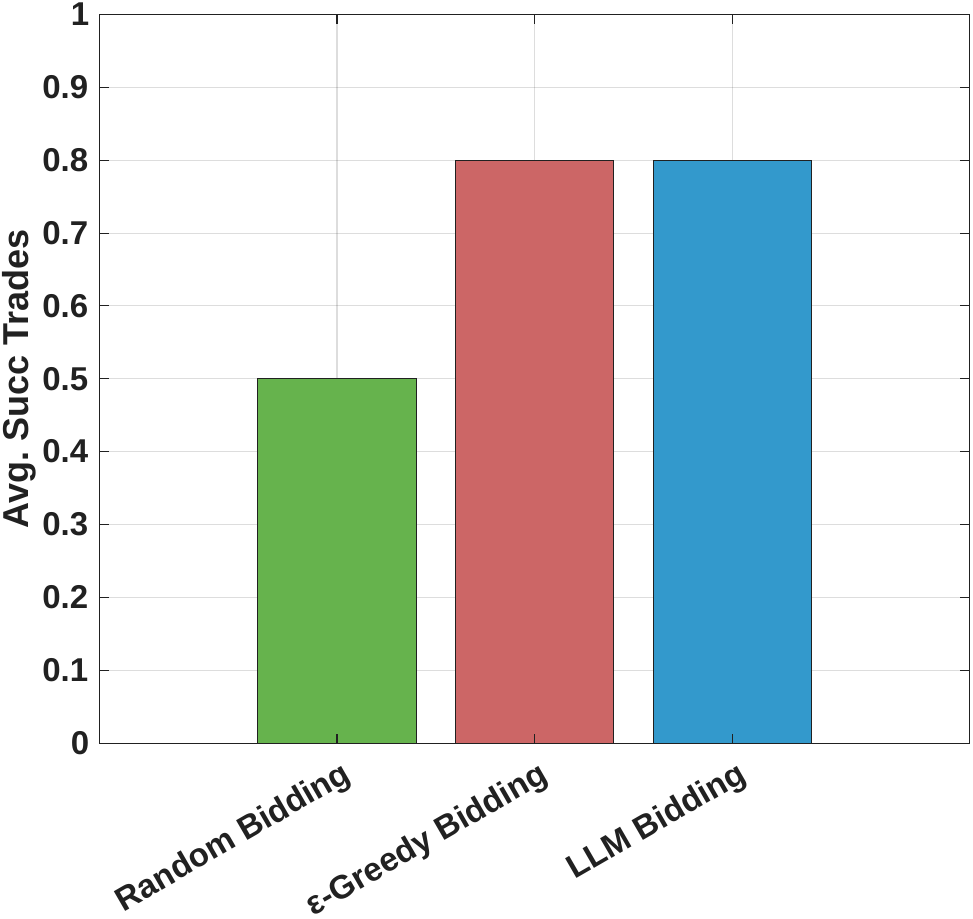}
         \caption{}
         \label{fig:avg_succ_trade}
     \end{subfigure}
     \begin{subfigure}[b]{0.24\textwidth}
         \centering         \includegraphics[width=\textwidth,height=3.1cm]{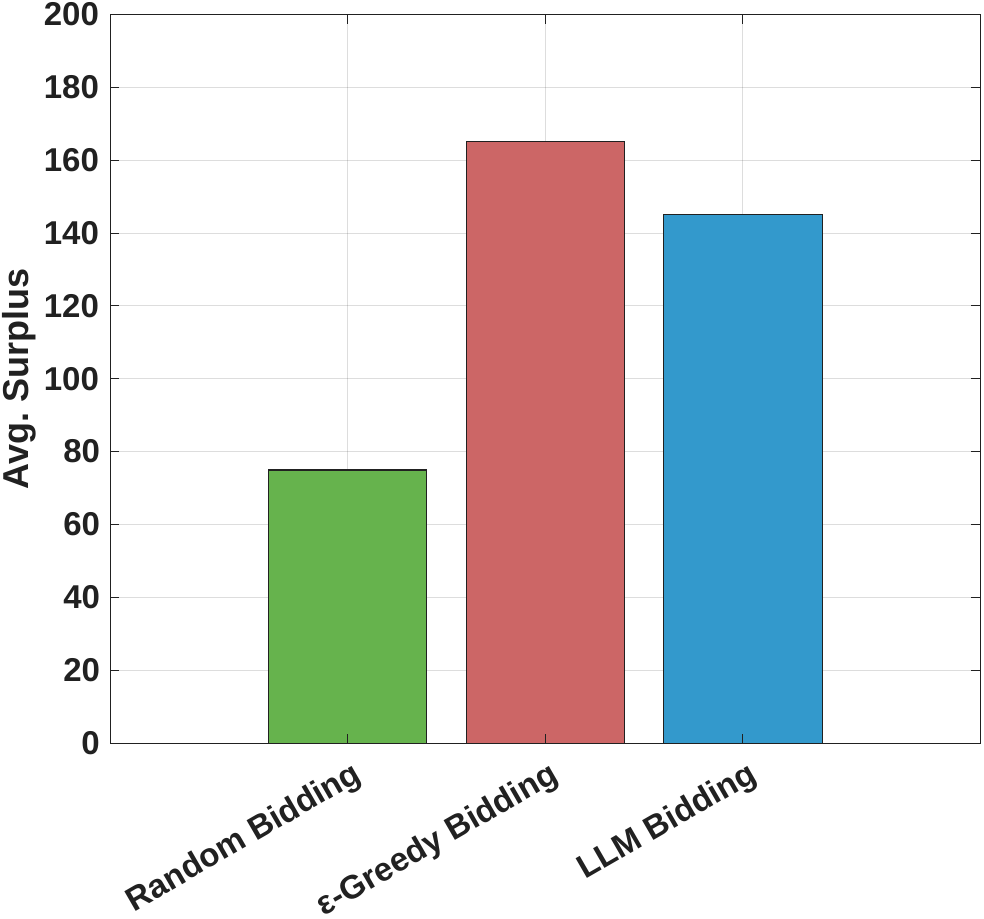}
         \caption{}
         \label{fig:accum_surplus}
     \end{subfigure}
     \caption{(a) Average number of successful trades, and (b) average agents' surplus for different bidding strategies.
        }        \label{fig:E2}
\end{figure}

\begin{figure}[ht] 
     \centering
     \begin{subfigure}[b]{0.24\textwidth}
         \centering         \includegraphics[width=\textwidth,height=3.1cm]{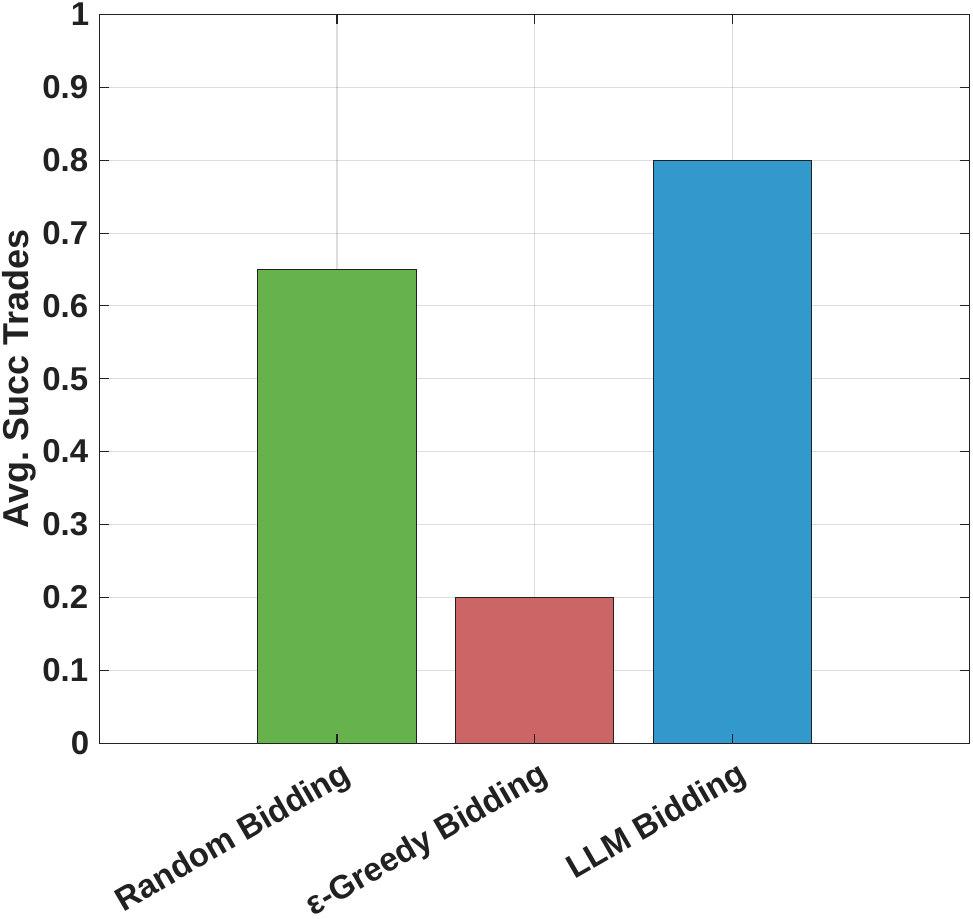}
         \caption{}
         \label{fig:avg_succ_trade_deploy}
     \end{subfigure}
     \begin{subfigure}[b]{0.24\textwidth}
         \centering         \includegraphics[width=\textwidth,height=3.1cm]{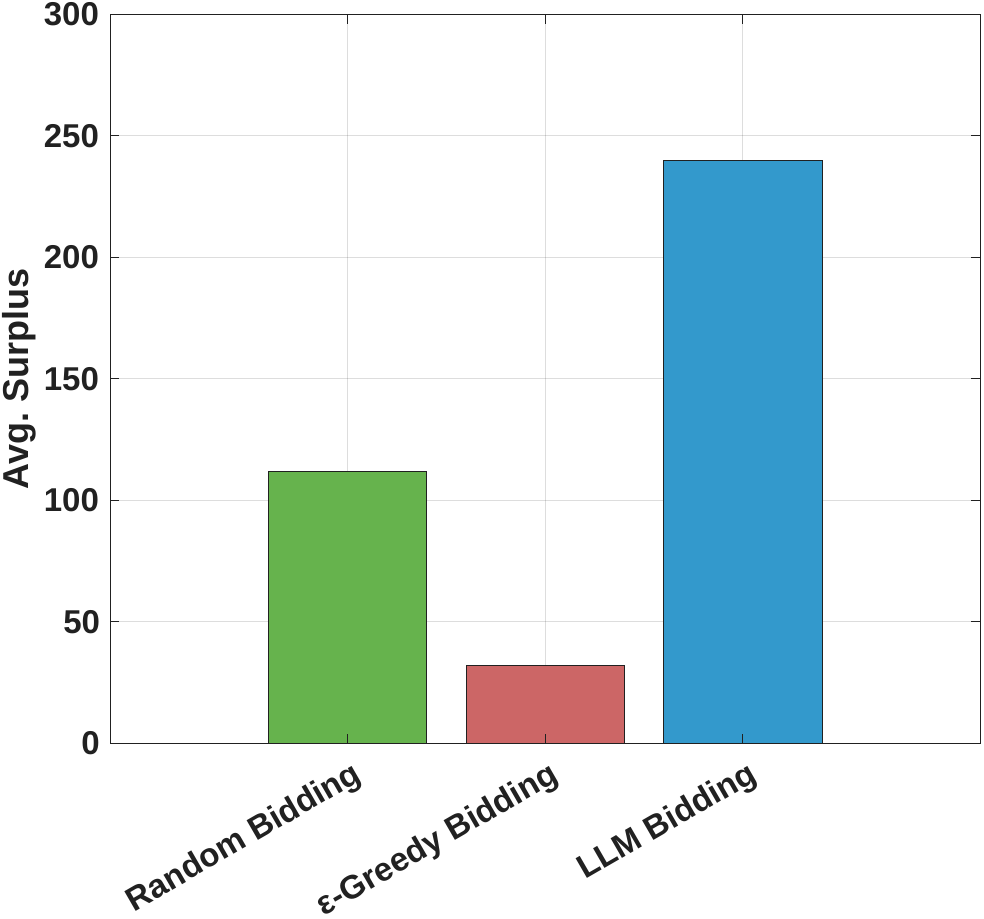}
         \caption{}
         \label{fig:accum_surplus_deploy}
     \end{subfigure}
     \caption{(a) Average number of successful trades, and (b) average agents' surplus for different bidding strategies, after market changes.
        }        \label{fig:E6}
\end{figure}

The results presented so far show that the proposed LLM-based bidding strategy and the $\varepsilon$-greedy MAB algorithm achieve comparable long-term performance after the latter converges. This naturally raises the question of whether the additional computational cost of LLM inference is justified.
We argue that the primary advantage of LLMs lies in their ability to reason from the current market context rather than relying on a policy learned for a specific environment. Consequently, LLM-based agents are expected to be more robust to distribution shifts after deployment, whereas the performance of learning-based methods may degrade until sufficient re-exploration or retraining has occurred.
To evaluate this capability, we introduce a distribution shift after deployment by increasing both the feed-in tariff and retail electricity price by $7$\,\textcent/kWh, resulting in $P_{\text{FIT}}=17$\,\textcent/kWh and $P_{\text{UR}}=27$\,\textcent/kWh. The bidding agents are then evaluated without any additional training or parameter updates.
Figure~\ref{fig:E2} presents the average number of successful trades and the average agent surplus under the original environment, while Figure~\ref{fig:E6} reports the corresponding results after the distribution shift. In both experiments, the agent population is evenly divided among the three bidding strategies, with one-third using random bidding, one-third using the $\varepsilon$-greedy algorithm, and one-third using the proposed LLM-based strategy.

While the $\varepsilon$-greedy and LLM strategies exhibit comparable performance under the original environment (Figure~\ref{fig:E2}), both in terms of the average number of successful trades and the average agent surplus, the performance of the $\varepsilon$-greedy algorithm degrades substantially after the distribution shift (Figure~\ref{fig:E6}). In fact, the performance of the random bidding strategy becomes better than the $\varepsilon$-greedy strategy. Unlike $\varepsilon$-greedy, which continues to favor price arms that accumulated high rewards in the original environment, the random strategy immediately shifts its sampling interval to the new price range and therefore remains competitive. Since the $\varepsilon$-greedy agent has not yet explored the newly optimal price arms, its bids are frequently outside the clearing region, resulting in fewer successful trades and lower surplus.
In contrast, the performance of the proposed LLM-based strategy remains largely unchanged following the distribution shift, supporting the hypothesis that reasoning-based agents are inherently more robust to changes in market conditions. By interpreting the updated market context, the LLM immediately generates bids within the new clearing region rather than rediscovering profitable price levels through exploration, which is required by statistical learning methods such as $\varepsilon$-greedy.

\section{Conclusion}
This paper investigated the integration of LLMs as strategic bidding agents in repeated P2P energy double auctions within smart grid distribution networks, benchmarked against multi-armed bandit-based and random bidding strategies under the $k$-double auction mechanism. The proposed LLM-based framework leverages contextual reasoning over historical auction outcomes to dynamically adapt bids across episodes, bypassing the statistical burn-in period that MAB-based agents inherently require before converging to productive price arms.
Simulation results demonstrated that the LLM strategy achieves higher cleared trading volume and surplus in the critical early deployment phase, a practically significant advantage corresponding to approximately 40 calendar days of suboptimal market performance under $\varepsilon$-greedy bidding. 
Notably, when market conditions deviate from those encountered during the learning phase, the $\varepsilon$-greedy agent suffers a marked performance degradation, as its converged price arms are no longer aligned with the new market environment and relearning from scratch is required. The LLM agent, by contrast, requires no such relearning as it reasons over the current market context at each episode and adapts its bidding strategy immediately, maintaining consistently high surplus and trade success rates regardless of the deployment environment.

However, the asymmetric market outcome in which sellers capture an increasing and disproportionate share of the available trading surplus at buyers' expense requires further investigation. Future work should examine whether this seller-side escalation behavior persists across alternative double auction mechanisms or if any fairness method can be used to achieve the theoretical Nash equilibrium. Finally, it is also worth confirming that other learning methods such as \ac{RL} and \ac{MARL} suffer from the same burn-in period experienced in MAB methods.

\bibliographystyle{IEEEtran}
\bibliography{ref}

\end{document}